# Tension-compression asymmetry in superelasticity of SrNi$_2$P$_2$ single crystals and the influence of low temperatures

Shuyang Xiao[1], Adrian Valadkhani[2], Sarshad Rommel[1], Paul C. Canfield[3], Mark Aindow[1], Roser Valentí[2], Seok-Woo Lee[1,*]

1. Department of Materials Science and Engineering & Institute of Materials Science, University of Connecticut, 25 King Hill Road, Unit 3136, Storrs CT  06269-3136, USA

2. Institute of Theoretical Physics, Goethe University, Frankfurt am Main, D-60438 Frankfurt am Main, Germany

3. Ames Laboratory & Department of Physics and Astronomy, Iowa State University, Ames IA  50011, USA

Corresponding Author:

Seok-Woo Lee

- Address: Department of Materials Science and Engineering & Institute of Materials Science, 25 King Hill Road, Unit 3136, Storrs, CT 06269-3136, USA

- Email: seok-woo.lee@uconn.edu

- Telephone: +1 (860) 486-8028

**Abstract**

ThCr$_2$Si$_2$-type intermetallic compounds are known to exhibit superelasticity associated with structural transitions through lattice collapse and expansion. These transitions occur via the formation and breaking of Si-type bonds, respectively, under uniaxial loading along the [0 0 1] direction. Unlike most ThCr$_2$Si$_2$-type intermetallic compounds, which have either an uncollapsed tetragonal structure or a collapsed tetragonal structure, SrNi$_2$P$_2$ possesses a third type of collapsed structured: a one-third orthorhombic structure, for which one expects the occurrence of unique structural transitions and superelastic behavior. In this study, uniaxial compression and tension tests were conducted on micron-sized SrNi$_2$P$_2$ single crystalline columns at room temperature, 200K, and 100K, to investigate the influence of loading direction and temperature on the superelasticity of SrNi$_2$P$_2$. Experimental data and density functional theory calculations revealed the presence of tension-compression asymmetry in the structural transitions and superelasticity, as well as an asymmetry in their temperature dependence, due to the opposite superelastic process associated with compression (forming P-P bonds) and tension (breaking P-P bonds). Additionally, following thermodynamics, the observations suggest that this asymmetric superelasticity could lead to an opposite elastocaloric effect between compression and tension, which could be beneficial potentially in obtaining large temperature changes compared to conventional superelastic solids that show the same elastocaloric effect regardless of loading direction. These results provide an important fundamental insight into the structural transitions, superelasticity processes, and potential elastocaloric effects in SrNi$_2$P$_2$.

# Graphical Abstract

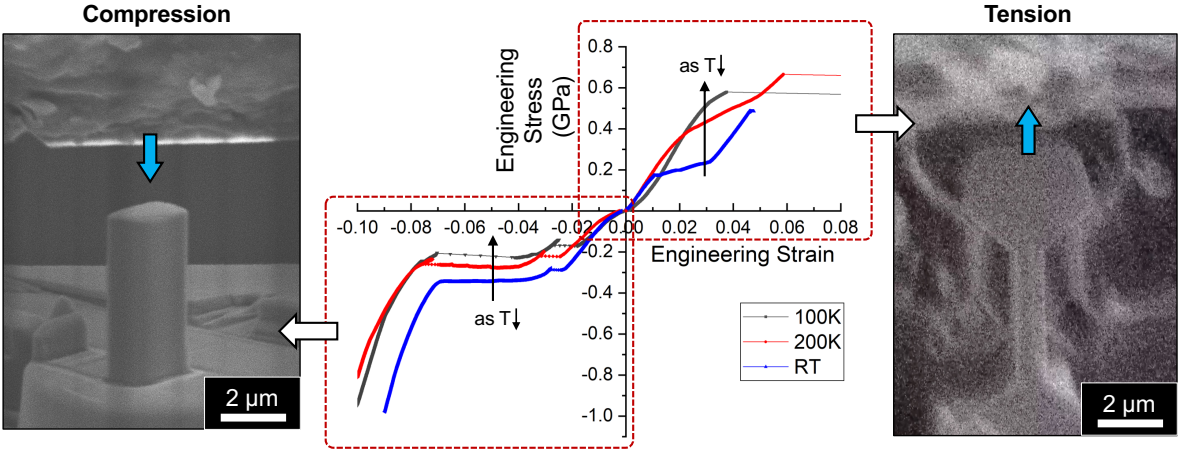

# 1. Introduction

Superelastic (or pseudo-elastic) deformation of crystalline solids usually occurs by a reversible structural transition that happens during the mechanical loading-unloading cycle [1]. Superelastic solids are an important class of structural materials because they can absorb mechanical energy and dissipate it as heat, offering an effective protection from mechanical deformation [2]. Shape memory alloys (SMA) are the most widely known superelastic solids, which undergo a reversible martensitic phase transformation [1], leading to a large elastic strain limit of between 3~15%. However, there have been challenges in finding SMAs that can exhibit both superelasticity at cryogenic temperatures and high fatigue resistance [3]. Spontaneous reversal of the shear transformation usually requires a large amount of thermal energy. Thus, only a limited number of SMAs exhibit superelasticity at cryogenic temperatures [4–6]. In addition, the shear deformation drives the glide of the mobile dislocations and changes the dislocation structures gradually over time. These evolved dislocation structures suppress the shear transformation process, degrade the superelastic performance, and finally reduce the fatigue lifetime [3]. Thus, achieving low temperature superelasticity and high microstructural stability have been important topic for the development of cryogenic superelastic materials, which could be used in engineering applications for outer space where temperature are typically very low [7].

Recently, $ThCr_2Si_2$-type intermetallic compounds have demonstrated a strong potential for both cryogenic superelasticity and exceptional microstructural stability [8–10]. Particularly, their cryogenic linear shape memory effect could be used to develop a linear actuator that can be used in extremely cold space environments [8]. Micropillar compression studies have demonstrated the presence of superelasticity in $CaFe_2As_2$ [8], $LaRu_2P_2$ [10], $CaKFe_4As_4$ [11], and $SrNi_2P_2$ [9] single crystals. Density functional theory (DFT) calculations showed that forming and breaking Si-Si-type bonds (As-As bonds for $CaFe_2As_2$ and $(CaK)_{0.5}Fe_2As_2$, and P-P bonds for $LaRu_2P_2$ and $SrNi_2P_2$) during compressive loading and unloading is the fundamental mechanism of superelasticity. Furthermore, cryogenic micro-compression tests revealed the presence of superelasticity even below 50K [8]. Breaking As-As or P-P bonds during unloading requires much less thermal energy than the reversal of shear phase transformations in SMAs, so the collapsed lattice can expand back to the original shape spontaneously at low temperatures. In addition, forming and breaking atomic bonds is a local, quantum-mechanical process, so the repeated structural transitions during cyclic loading do not cause any significant changes in microstructure. Thus, $ThCr_2Si_2$-type intermetallic compounds have also demonstrated superior microstructural stability [9].

Among $ThCr_2Si_2$-type intermetallic compounds, it was found that a $SrNi_2P_2$ single crystal is a unique case because of the coexistence of two different crystal structures under ambient conditions: third collapsed orthorhombic (tcO) superstructure (~85 vol.%) and uncollapsed tetragonal (ucT) structure (~15 vol.%) [9, 12]. One third of the P atoms in the tcO superstructure are bonded to one another, while the ucT structure has no P-P bonds. Due to the co-existence of two different crystal structures, a two-step lattice collapse and expansion was observed under compressive loading and unloading along the [0 0 1] direction [9]. During compression, the ucT structure transforms to the tcO superstructure (the first collapse), and then the entire sample volume with the tcO superstructure transforms to the collapsed tetragonal structure (cT) (the second collapse) under further compression. During unloading, the cT structure firstly reverts to the tcO superstructure (the first expansion), and then the tcO superstructure transforms partially to the ucT structure (the second expansion). Through this unique double lattice collapse and expansion, $SrNi_2P_2$ exhibits a giant elastic strain limit of ~14%. In addition, cryogenic micro-compression tests confirmed that this double lattice collapse happens even at 100K, confirming that superelasticity occurs even at low temperatures.

However, one major drawback in superelasticity of $ThCr_2Si_2$-type intermetallic compounds is that it is difficult to observe the structural transition under compressive and tensile loading in the same material. If $ThCr_2Si_2$-type intermetallic compounds initially exhibit the ucT structure that has no Si-Si type bonds, the superelasticity occurs only under compression through forming Si-Si type bonds. Conversely, if $ThCr_2Si_2$-type intermetallic compounds have initially the cT structure where all Si-type atoms are bonded, the superelasticity occurs only under tension through breaking Si-Si type bonds. Thus, it appears to be impossible to observe compressive and tensile superelasticity in a single system, because the initial crystal structure must be either the ucT structure or the cT structure. In contrast, conventional SMAs, for instance, NiTi, undergo the energetically equivalent reversible martensitic phase transformation process regardless of loading direction [13]. The uni-directional transition of $ThCr_2Si_2$-type compounds could limit their applicability in linear actuators that require operation under a wide range of stress states. In addition, superelasticity is known to induce the elastocaloric effect, in which a material can heat up or cool down through deformation. The uni-directional transition could also limit the range of temperature changes compared to a two-directional transition.

Although most $ThCr_2Si_2$-type intermetallic compounds do not exhibit superelasticity under compressive and tensile loading at the same time, we hypothesized that $SrNi_2P_2$ could exhibit superelasticity under both compressive and tensile loading due to the unique presence of the tcO superstructure, where only 1/3 of P-P pairs are bonded. Such structures have not been found in most $ThCr_2Si_2$-type intermetallic compounds, and $SrNi_2P_2$ is the only system in which the presence of tcO superstructure has been reported. We already confirmed that under compression, the tcO superstructure becomes the cT structure through the formation of P-P bonds between the unbonded 1/3 of P-P pairs [9]. Under tension, it could be possible that the tcO superstructure becomes the ucT structure if the tensile loading breaks P-P bonds (1/3 of P-P bonded pairs). Furthermore, if tensile superelasticity does occur, it is also important to investigate the effect of low temperatures on the structural transition and superelastic behavior under tension, and to compare the results with its effect under compression. Since forming and breaking atomic bonds is highly sensitive to temperature, superelasticity under both compression and tension could be affected significantly by temperature changes. Our previous work confirmed that under compression, the structural transition (lattice collapse via forming P-P bonds) occurs at a lower stress at lower temperatures [9]. $SrNi_2P_2$ under tensile loading could exhibit a different temperature dependence of superelasticity because the structural transition under tension is a completely opposite process (lattice expansion via breaking P-P bonds).

In this present study, therefore, we conducted micro-compression and micro-tensile tests on a $SrNi_2P_2$ single crystal along the [0 0 1] direction at various low temperatures and investigated how the tensile superelasticity and lattice expansion process are influenced by the temperature change. The results showed that $SrNi_2P_2$ single crystals indeed exhibit superelasticity (~5% in strain), even under tensile loading and unloading. To our knowledge, this is the first report that demonstrates the presence of both compressive and tensile superelasticity from a single $ThCr_2Si_2$-type intermetallic compound. Unlike the compressive superelasticity with a two-step lattice collapse and expansion [9], however, tensile superelasticity occurs through one-step lattice expansion and collapse. Also, the structural transition occurs at a higher stress at lower temperatures. Thus, $SrNi_2P_2$ exhibits a strong tension-compression asymmetry in the structural transition process and its temperature dependence. Interestingly, it was found that the tension-compression asymmetry leads to the opposite elastocaloric effect between compression and tension unlike conventional SMAs that show the same elastocaloric effect regardless of loading direction. Tension-compression asymmetry will be discussed in terms of the role of P-P distance and thermal vibration effects. Elastocaloric effects will be explained based on a thermodynamic calculation with the Clausius-Clapeyron equation [14]. These experimental and computational results provide an important fundamental insight into the unique superelasticity mechanisms of $SrNi_2P_2$ and also demonstrate that $SrNi_2P_2$ holds great promise as an excellent superelastic and elastocaloric material.

## 2. Experimental Method

### 2.1 Single crystal growth

Single crystals of $SrNi_2P_2$ were grown from a quaternary melt with an initial composition of $Sr_{1.3}Ni_2P_{2.3}Sn_{16}$. High purity elements were placed into a 2ml, fritted alumina crucible set [15], sealed into an amorphous silica ampoule under a partial atmosphere of high purity Ar and placed in a vented box furnace [16]. The furnace was then heated to 600 °C over 4 h, held at 600 °C for an additional 4 h, heated to 1150 °C over 5 h, held at 1150 °C for 24 h, and then cooled to 650 °C over 250 h. At 650 °C, the ampoule was removed from the furnace and placed in a centrifuge for decanting of the excess melt from the $SrNi_2P_2$, crystalline phase. Plate-like single crystals with dimensions of ~ 3x3x0.2 mm$^3$ were common with some crystals reaching up to 6 times that volume. A more detailed description of the single crystal growth process can be found in Ref. 9 and 12.

### 2.2. Fabrication of micro-mechanical testing specimens

An FEI Helios Nanolab 460F1 Ga+ ion dual beam focused-ion-beam scanning electron microscope (FIB-SEM) was used to fabricate square cross-section compression micropillars with 2μm width and 6μm height (Fig.1 a) as well as dog-bone shaped micro-tensile specimens with ~2μm width, ~2μm thickness, and 10μm gauge length (Fig.1 b). Gallium ion beam currents from 300pA to 10pA were used from initial to final thinning at an operation voltage of 30kV. Because the typical thickness (10~20nm) of the FIB damage layer is much less than the width of the micropillar (~2μm), FIB damage effects on mechanical data are expected to be negligible. The effect of sample size on the mechanical properties is expected to be negligible, due to the much smaller length scale of the phase transformation(s) (unit cells) compared with the length scale of micropillars (a few micrometers).

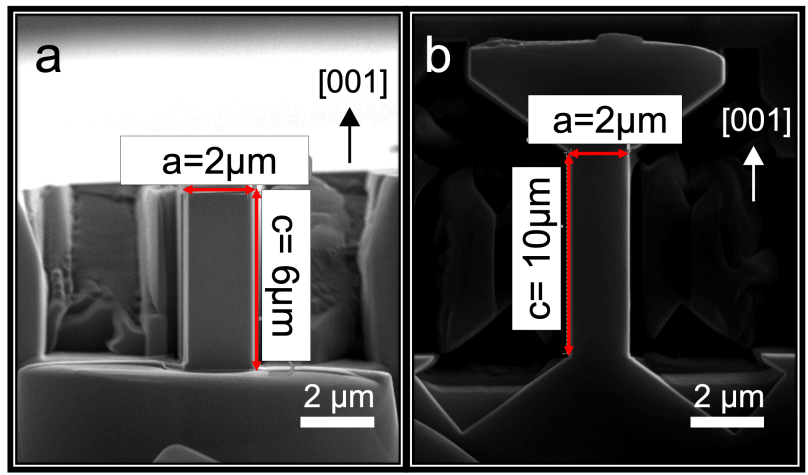

**Figure 1.** SEM images of micropillars used for a) micro-compression test b) micro-tensile test.

### 2.3. In-situ nanomechanical testing

Micromechanical compression and tension tests along the [0 0 1] direction (c-axis of the crystal) were performed under ultra-high vacuum conditions (<$10^{-4}$Pa) at different temperatures (room temperature, 200K, and 100K) using a NanoFlip$^{TM}$ stage (KLA, TN, USA), installed in a field-emission gun JEOL 6335F scanning electron microscope (JEOL, Japan). The top and bottom surfaces of the solution grown crystal are relatively large and flat {001} planes. This allows for great alignment between the micropillar axis and the loading direction and minimizes any misorientation between the micropillar and the diamond tip. The excellent alignment between micropillar and diamond tip was also confirmed by the presence of a stress plateau in compressive stress-strain data. A misalignment between the micropillar and the diamond tip would result in a gradual slope in the stress-strain curve rather than a stress plateau. A nominal displacement rate of 10nm/s was used, and this corresponds to an engineering strain rate of around $0.0016s^{-1}$. Low temperature tests were performed by using a customized cryostat (Janis Research, MA, USA) and liquid helium. For all cryogenic tests in this study, thermal equilibration between the tip and sample was maintained, so the thermal drift was always below 0.5nm/s. Recorded videos were also used to confirm visually the accuracy of our strain measurements. A more detailed description of the cryogenic mechanical testing is also available elsewere [11]. Also, note that most compressive stress-strain curves in this paper use positive values of compressive stress when they are plotted separately. If they are plotted together with tensile stress-strain curves in a single graph, negative values of compressive stress were used by convention.

## 2.4. Transmission Electron Microscopy (TEM)

The cross-sectional TEM specimens were also prepared in the Helios Nanolab 460F1 FIB-SEM using the lift-out method. In this method, a 2µm thick layer of Pt was first deposited on the sample surface to minimize ion-beam damage to the sample during this procedure. A lamella containing a cross-section of the sample surface was then cut free from the sample using the ion beam and transferred onto a Cu omni-grid using an EasyLift micromanipulator needle. This lamella was then thinned using the ion beam to electron transparency (~100 nm in thickness) using iteratively lower ion beam currents at 30kV, to a final current of 80 pA. These cross-sectional TEM specimens were examined through TEM, selected area electron diffraction (SAED) and energy dispersive X-ray spectroscopy (EDXS) analysis using a Titan Themis probe-corrected scanning transmission electron microscopy (STEM) operated at an accelerating voltage of 300 kV. X-ray maps revealing the spatial distribution of each element were generated from EDXS spectrum images obtained from the specimens; these maps were extracted using the intensities of the K-lines for the relevant elements at each pixel.

## 2.5. Density Functional Theory calculations

All calculations were performed within ab *initio* DFT [17, 18]. For this we used the Vienna ab initio simulation package (VASP - version 6.3.0) [19, 20]. The planewave basis set and projector augmented wave (PAW) [21, 22] was employed as implemented in VASP. The exchange-correlation contribution was treated using the PBEsol functional [23]. The energy cutoff was set to 600 eV. Each calculation consists of a two-step procedure [9]. (i) First, the $1 \times 1 \times 1$ structure was relaxed using a $9 \times 9 \times 4$ grid and the $1 \times 3 \times 1$ structure using a $9 \times 3 \times 4$ grid. For both structures a force convergence lower than $10^{-3}$ eV/Å for VASP was used. (ii) In a second step the energy was calculated using much denser grids. A $21 \times 21 \times 8$ grid was chosen for the $1 \times 1 \times 1$ structure and a $21 \times 7 \times 8$ grid for the $1 \times 3 \times 1$ structure. By comparing both energies we extracted the ground state (or lower lying state) at a given strain according to DFT. The resulting changes of the *a* and *b* parameters were small enough to approximate the *k*-mesh of the $1 \times 3 \times 1$ by using $1 \times 1 \times 1$ where we divide the maximum of $k_y$ by three. This two-step procedure was repeated for each value of strain.

The strain value is given with respect to the ground state which resembles the experiment at ambient pressure of the given structure. Strain was simulated by adding a value ε to the ground state c parameter as c = $c_0(1 + \varepsilon)$. The sign of $\varepsilon$ determines the type of strain, where positive (negative) values represent tensile (compressive) strain. To obtain the pure uni-axial stress state, the lateral stress components were canceled by adjusting the lateral dimensions. For a given axial deformation at every step, lateral stresses ($\sigma_{xx}$ and $\sigma_{yy}$) were relaxed iteratively by adjusting the lateral dimensions until both $\sigma_{xx}$ and $\sigma_{yy}$ become zero. Different exchange-correlation functionals like PBE [24] or LDA [25-27] were used for comparison. However, it turns out that the PBEsol functional is the best one to describe the tcO superstructure. As a crosscheck QuantumEspresso [28, 29] (version 6.8) was used and supports the results from VASP very well.

## 3. Experimental Results

### 3.1 Uniaxial compression tests of SrNi$_2$P$_2$ micropillars along the c-axis

The room temperature compressive stress-strain curve with twenty loading-unloading cycles shows that there are five distinct stages of deformation, which include two plateau regions and three elastic regions, in both the loading and unloading curves (Fig. 2a). First, the initial linear elastic deformation (Stage I) occurs. Then, the first plateau (Stage II) appears at ~0.28GPa. Here, the strain range of the first plateau is only ~0.5%, which is relatively small. Then, after the second elastic deformation (Stage III), the second plateau (Stage IV) appears at ~0.33GPa with a strain range of ~3.5%, which is much larger than the first plateau. Finally, there is a third elastic deformation region (Stage V). If unloading begins before fracture happens, the five stages of deformation can be reversed fully. Two plateau regions appear at ~0.22GPa (Stage IV') and ~0.17GPa (Stage II') during unloading. The hysteresis exists due to the different stress levels for the structural transitions between loading and unloading. The shape of the stress-strain curve of SrNi$_2$P$_2$ resembles those of conventional SMAs, but SrNi$_2$P$_2$ shows two plateaus while most SMAs only show only one plateau. These data are consistent with our previous result [9]. Cyclic stress-strain curves also confirmed that these five stages of deformation and their reversed deformation always occur in both loading and unloading curves, respectively (See also Supplementary Video). Thus, SrNi$_2$P$_2$ is truly superelastic. In our previous work, a single loading test was also conducted until fracture occurred, and the average elastic strain limit of our sample was around 14%, which is significantly larger than that of most crystalline solids [9]. This giant elastic limit enables SrNi$_2$P$_2$ to have an ultrahigh modulus of resilience (146 MJ/m$^3$).

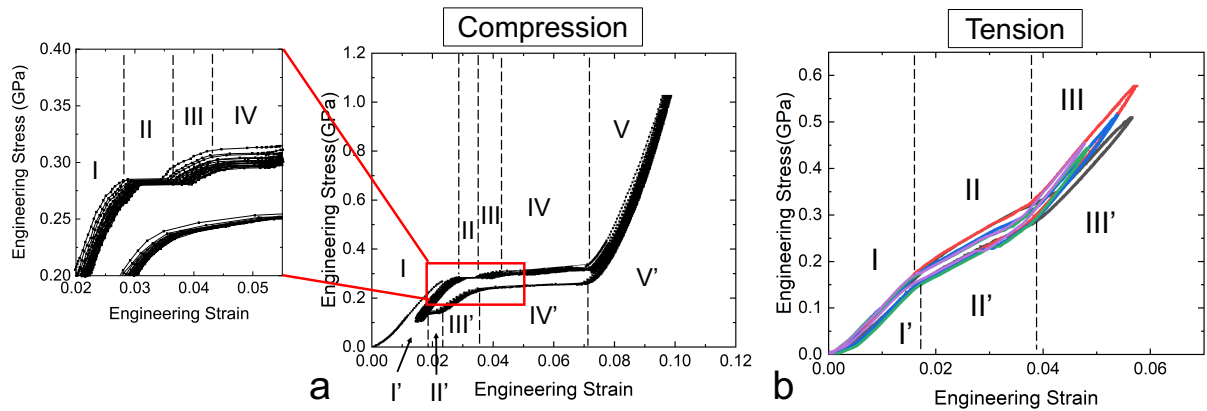

**Figure 2.** Room temperature stress-strain curves of (a) 20 cycles of compression testing and (b) 5 cycles of tension testing. The left graph in Fig. 2a shows the magnified view of the first four stages of compressive deformation.

### 3.2 Uniaxial tension tests of SrNi$_2$P$_2$ micro specimens along the c-axis

Uniaxial micro-tensile tests on $SrNi_2P_2$ micro-dog-bone specimens were conducted along the [0 0 1] direction (Fig. 2b) (See also Supplementary Video). Interestingly, loading-unloading tests showed that the shapes of the tensile stress-strain curves are different from those of the compressive stress-strain curves. Instead of two plateaus, only one region with a lower slope (Stage II) appears from 0.15GPa (onset) to 0.3GPa (offset) with a strain range of 2% (Fig. 2b). With two elastic regions (Stages I and III), thus, there are three stages of deformation under tension, which are reversible. Due to the possible premature fatigue failure and the experimental difficulties related to sample fabrication, sample alignment, and sample gripping, it was difficult to perform a programmed cyclic tensile test over a large number of cycles. Thus, an individual tensile test was done five times instead. Nevertheless, five-cycle tensile stress-strain curves confirm that the three stages of tensile deformation are fully reversible. Also, no obvious permanent damage was observed after loading-unloading tensile tests, indicating that $SrNi_2P_2$ is superelastic even under tension. Note that this is the first time that tensile superelasticity has been observed in $ThCr_2Si_2$-type intermetallic compounds. Once the tensile loading is applied further, fracture always occurs on the (0 0 1) plane (Fig. 3).

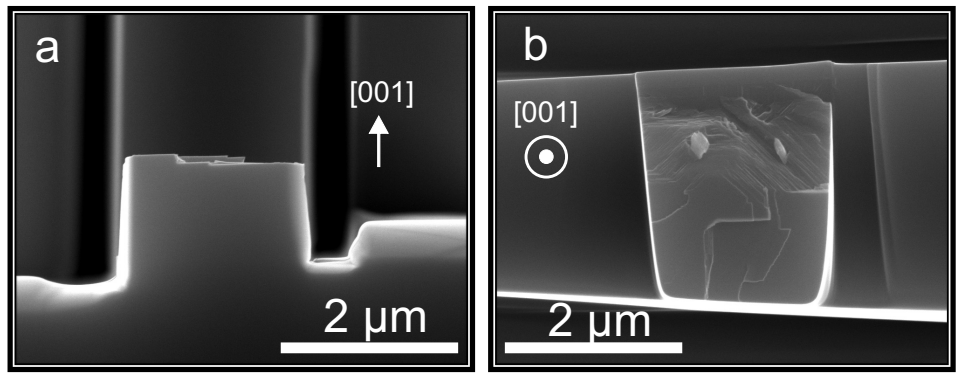

**Figure 3.** SEM images of fractured tension micropillar. (a) side view (b) top view

**3.3 Stress-strain data at different cryogenic temperatures**

Uniaxial compression and tension tests were performed along the [0 0 1] direction at room temperature, 200K, and 100K. Note that compression data were obtained from our previous work [9]. Compressive stress-strain data show that the five stages of deformation still exist even at 200K and 100K. Notably, the two plateau regions in each stress-strain curve appear at lower compressive stresses at lower temperatures, respectively (Fig. 4a). On the other hand, tensile stress-strain data show that the low slope region is initiated at higher tensile stresses at lower temperatures. However, tensile fracture occurs right after the end of the transition (200K) or in the middle of the transition (100K) (Fig. 4b). Under tension, therefore, if the onset stress of the structural transition becomes higher than the fracture strength, the transition cannot be observed. The combination of the stress-strain data of both compression and tension shows clearly that $SrNi_2P_2$ exhibits an asymmetrical stress-strain curve (Fig. 4c). Here, to combine compression and tension stress-strain curves in a single plot, Compressive stresses are presented with negative values, and tensile stresses are presented with positive values by convention. Under this sign convention, when temperature decreases, the transition stresses of both compressive and tensile stress-strain curves of $SrNi_2P_2$ move upwards. This result indicates that under compression, less compressive stress is required to induce a structural transition as the temperature decreases. Under tension, however, a larger tensile stress is required to induce a structural transition.

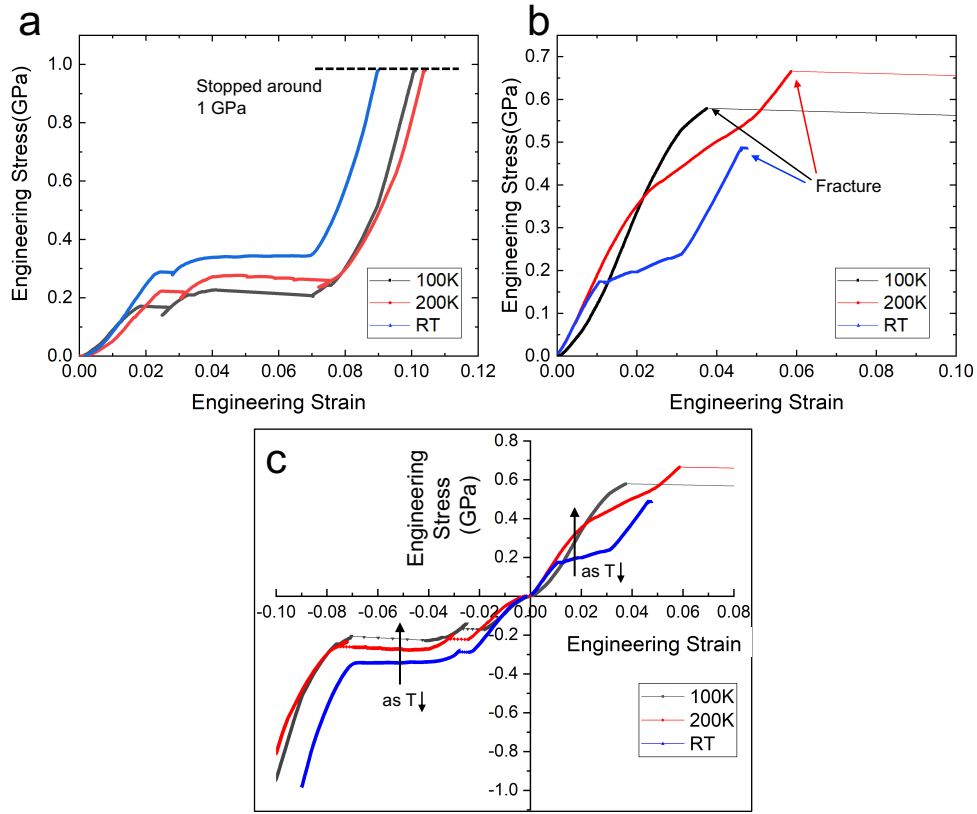

**Figure 4.** Stress-strain curves of (a) Compression test stopped around 1 GPa at 100K, 200K and room temperature [9] (Reprinted with permission from Ref. [9]. Copyright 2021 American Chemical Society). Here, the tests were stopped around 1GPa because this is enough to show how the structural transition occurs at different temperatures. (b) Tension test until fracture at 100K, 200K and room temperature. Tension tests had to be done until fracture occurs because fracture sometimes occurs before the structural transition is finished. (c) Combining compression and tension tests. Note that to combine compression and tension stress-strain curves in a single plot, compressive stresses are presented with negative values, and tensile stresses are presented with positive values by convention.

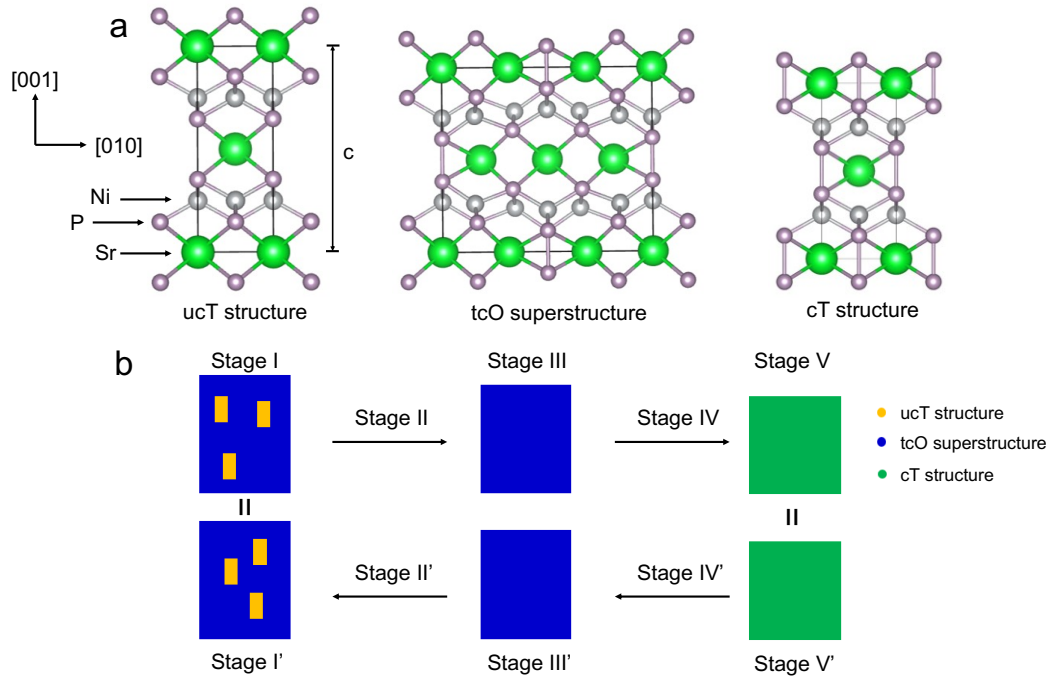

**Figure 5.** Phase transformation between the tcO superstructure and the ucT structure during loading and unloading of compression test. (a) Crystal structures of the ucT structure, the tcO superstructure and the cT structure (b) Schematic diagram of structural transition sequence during compressive loading and unloading. Here, note that the locations of ucT structure in Stage I' is not necessarily the same to those in Stage I because the T phases in Stage I' could be formed due to thermal fluctuation during Stage II'.

## 4. Discussion

### 4.1 Double lattice collapse and expansion processes under compression

In previous work, X-ray diffraction and electron diffraction data confirmed that $SrNi_2P_2$ exhibits two different crystal structures, the tcO superstructure and the ucT structure, under the zero-stress state and at room temperature (Fig. 5a) [9]. X-ray diffraction patterns clearly showed the combined diffraction data of the tcO superstructure and the ucT structure. Although these two structures are indistinguishable through electron diffraction due to their structural coherency, we obtained two different high-resolution TEM images that correspond to the the tcO and ucT structures. Therefore, the tcO and ucT structures coexist at zero-stress state and at room temperature. It is interesting that the tcO phase is the only available superstructure, and other superstructures with a large unit cell size (*e.g.,* fifth collapsed orthorhombic) do not exist. It is worthwhile to pursue an in-depth theoretical study to understand the unique atomic arrangement of the tcO superstructure. Also, DFT calculations provided the lattice parameters of the tcO superstructure, the ucT structure, and the cT structure [9]. The tcO superstructure has a smaller c lattice parameter (10.432Å) and a larger volume fraction (85%) than the ucT structure, where the c lattice parameter is 10.677Å and the volume fraction is 15%. TEM data also confirmed that these two phases are coherently connected [9]. Due to the strain compatibility, the tcO superstructure (blue phase in Fig. 5b) could be slightly under tension along the [0 0 1] direction, and the ucT structure (yellow phase in Fig. 5b) could be slightly under compression. In this case, if a compressive stress is applied along the [0 0 1] direction, the tcO superstructure becomes energetically more stable because the tcO superstructure relieves the initial residual tensile stress, but the ucT structure will be further compressed in addition to the initial residual compression. As a result, the compressed ucT structure will be transformed into the tcO superstructure by forming P-P bonds between one third of P atoms in the ucT structure. The transition from the ucT structure to the tcO superstructure corresponds to the first small plateau (Stage II in Fig. 2a) in the compressive stress-strain curve. The strain range of the first plateau must be small (0.5%) due to the small volume fraction (15%) of the ucT structure [9]. At the end of the first plateau, thus, the entire volume becomes the tcO superstructure, and its elastic deformation corresponds to the second elastic loading (Stage III in Fig. 2a). Then, the entire specimen with the tcO superstructure transforms to the cT structure (green phase in Fig.5 b), which has an even shorter c lattice parameter (9.761Å) than the tcO superstructure, due to the formation of P-P bonds between unbonded P atoms in the tcO superstructure (c=10.432Å). The strain range of the second plateau (3.5%) is much larger than that of the first one (0.5%) because the entire volume (100%) of the specimen is transformed at this time.

The DFT stress-strain curves of each structure show that the ucT structure requires a lower critical stress for lattice collapse than the tcO superstructure (Fig. 6a). Broken line arrows indicate the plateau region of the structural transition under constant loading condition, which corresponds to the experimental loading condition. Because our nanoindenter is a load-controlled machine, the load is maintained when the strain increases suddenly during the structural transition, leading to the plateau in experimental stress-strain data. Note that in a real experimental case, the immediate collapse from the ucT structure (10.677Å) to the cT structure (T→cT) would not occur because this process causes a large increase in energy due to the large c lattice mismatch between the tcO superstructure (10.432Å) and the cT structure (9.761Å). Thus, the transition from the ucT structure to the tcO superstructure is most likely to occur first by forming P-P bonds from only one third of P atoms. Then, after this intermediate transition, further compression collapses the entire volume with the tcO superstructure, leading to the full transition to the cT structure (O→cT marked by the broken black arrow in Fig. 6a). In this case, the two different P-P distances in the tcO superstructure converge into a single and short P-P distance in the cT structure, indicating that all P atoms form P-P bonds (Fig. 6b). Electron density distribution in the tcO superstructure under compression also confirms the significant increase in electron density between initially unbonded P atoms (marked by red arrows in Fig. 6c), indicating that the lattice collapse occurs via the formation of P-P bonds. This bonding process is similar to the formation of As-As bond in $CaFe_2As_2$ under compression along the [0 0 1] direction [8].

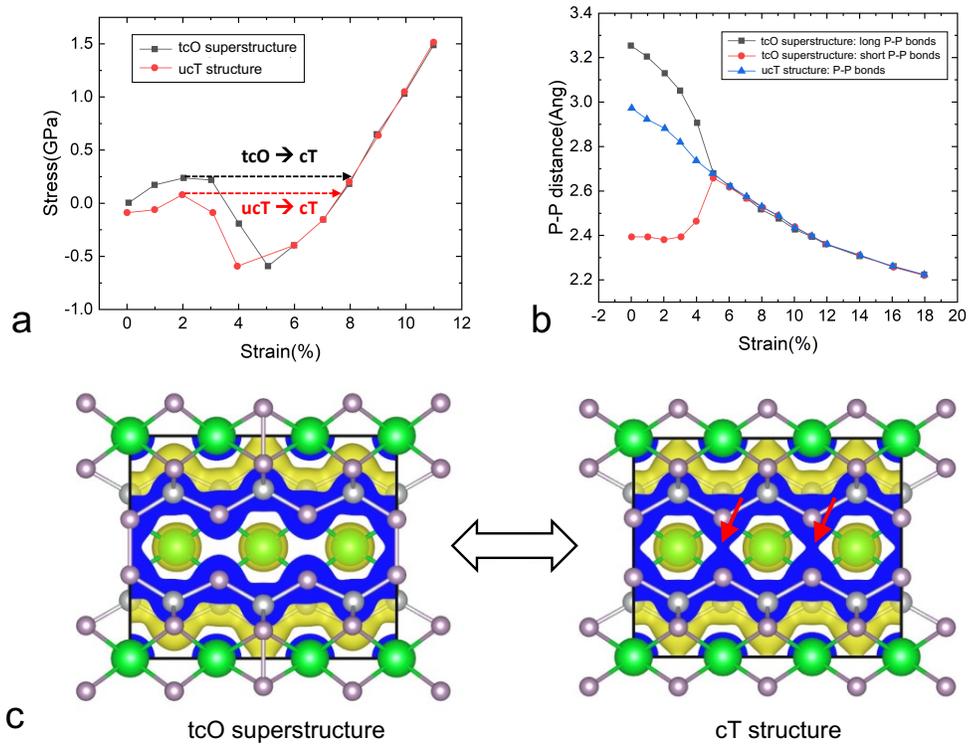

**Figure 6.** DFT calculation data of compression test. (a) Stress vs. Strain curve. A broken line arrow indicates the transition under load control. (b) P-P distance vs. Strain curve (c) Electron density before and after structural transition of the tcO superstructure. Red arrows indicate the location where formation or breaking P-P bonds occurs.

Upon unloading, the cT structure could become the tcO superstructure (Fig. 5b). This transition corresponds to the first large plateau (Stage IV') in the unloading curve. Further unloading induces the phase separation, so a fraction (~15%) of the tcO superstructure is transformed into the ucT structure. This transition corresponds to the second small plateau (Stage II'). We believe that the fraction of the recovered ucT structure must be the same as its original fraction prior to mechanical testing because the subsequent loading always produces identical stress-stress curves, which indicates that there is essentially the same microstructure at the beginning of each cycle. Before the final transition (Stage II') during unloading, the entire volume of the SrNi$_2$P$_2$ exhibits the tcO superstructure. Based on TEM data and our recent work [30], there are no structural defects or chemical inhomogeneities that could assist the creation of the ucT structure (Fig. 7). If chemical inhomogeneities produce the lattice distortion that creates the residual tensile stress along the [001] direction, the ucT structure could be preferentially nucleated at this region. However, we were not able to see any visible chemical inhomogeneities at the length scales of our TEM measurement, so we assume that our sample is chemically uniform and there is no structurally preferential nucleation site of the ucT structure. That is, the nucleation of the ucT structure could be a thermally-activated process, resulting in nucleation at random locations. However, the total volume fraction of the ucT phase must be constrained by thermodynamics. This means that the size, shape, and distribution of the ucT structure after the loading-unloading cycles could be different from those of the initial ucT structure before mechanical testing, but the total volume fraction (~15%) of the ucT structure must be maintained. The dynamics of ucT structure formation during unloading could be confirmed by performing synchrotron X-ray diffraction experiments with high-spatial resolution, but this may require a larger crystal and a higher force instrument. Such experiments are planned as part of a future study.

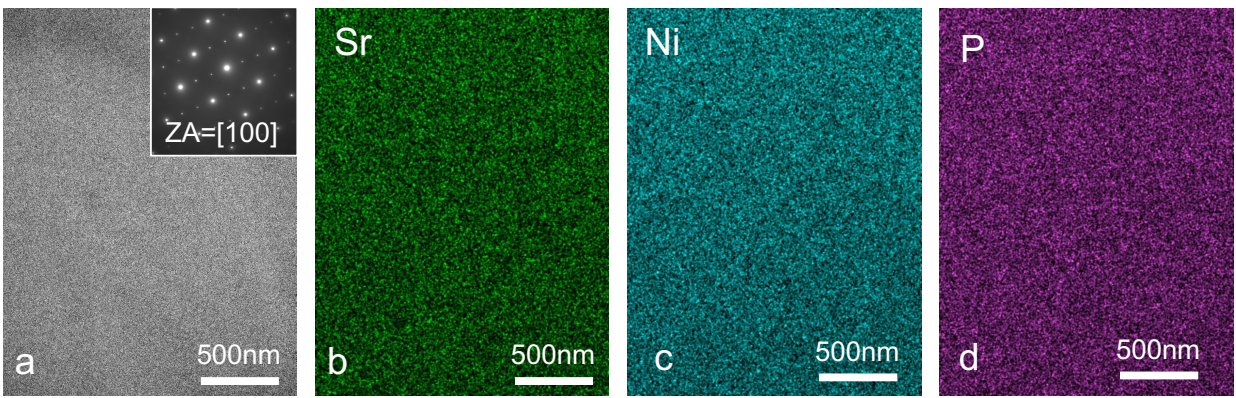

**Figure 7.** (a) Bright-field TEM images for [001] zone axis. The inset shows the diffraction pattern; Energy-dispersive X-ray spectroscopy map of (b) Sr, (c) Ni, and (d) P.

**4.2 Lattice expansion and collapse processes under tension**

As described in the previous section, the tcO superstructure is under residual tension, and the ucT structure is under residual compression with no applied stress due to the lattice mismatch. Upon tension, the tcO superstructure will experience more tension in addition to the initial residual tension, but the ucT structure will be able to relax the residual compression. Then, the enhanced tensile stress will break P-P bonds in the tcO superstructure, leading to a transition to the ucT structure, which has no P-P bonds (Fig. 8). Further tension will lead to the second elastic loading followed by fracture on the (0 0 1) plane (Fig. 3). The weak binding between (0 0 1) Th-type and Si-type atomic layers have been observed in $CaFe_2As_2$ (in this case, Ca and As layers), which could cause easy fracture on the (0 0 1) plane [31] and micaceous plasticity in (0 0 1)<1 0 0> slip systems [32]. Thus, the fracture could occur between the (0 0 1) Sr and P layers due to the weak layer bonding. Then, the entire volume will exhibit the ucT structure. It is worthwhile to note that the simultaneous presence of compression and tensile superelasticity is directly related to the presence of the tcO superstructure; this does not exist in most $ThCr_2Si_2$-structured intermetallic compounds, which are in either ucT or cT state. Partially bonded tcO superstructure can undergo not only the tensile-induced phase transition from the tcO to ucT structures by breaking P-P bonds but also compression-induced phase transition from the ucT to tcO structures by forming P-P bonds.

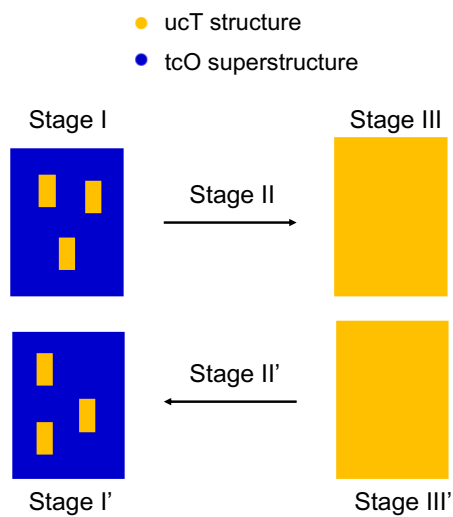

**Figure 8.** Schematic diagram structural sequence during tensile loading and unloading.

DFT calculations were also conducted to investigate the change in structural state. Tensile stress on the tcO superstructure leads to a sudden change in stress-strain (Fig. 9a), the P-P distance as a function of strain (Fig. 9b), and electron density distribution between P atoms (Fig. 9c). The DFT results show a sudden decrease in electron density between P atoms in the tcO superstructure under tension, implying that the P-P bonds are broken by a tensile loading. The ucT structure does not undergo any transition because it has no P-P bonds. Thus, fracture should occur under loading. One difference between the DFT and the experimental stress-strain curves is the absence of a plateau in the experimental curve; this latter curve has an inclined region (Stage II in Fig. 2b) during the structural transition. This difference could result mainly from the different shapes of the energy-strain curves. The compressive energy-strain curve has a deep energy well while the tensile energy-strain curve has a shallow energy well that could be removed relatively easily at a finite temperature. A more detailed discussion on the inclined transition curve under tension is available in Supplementary Materials.

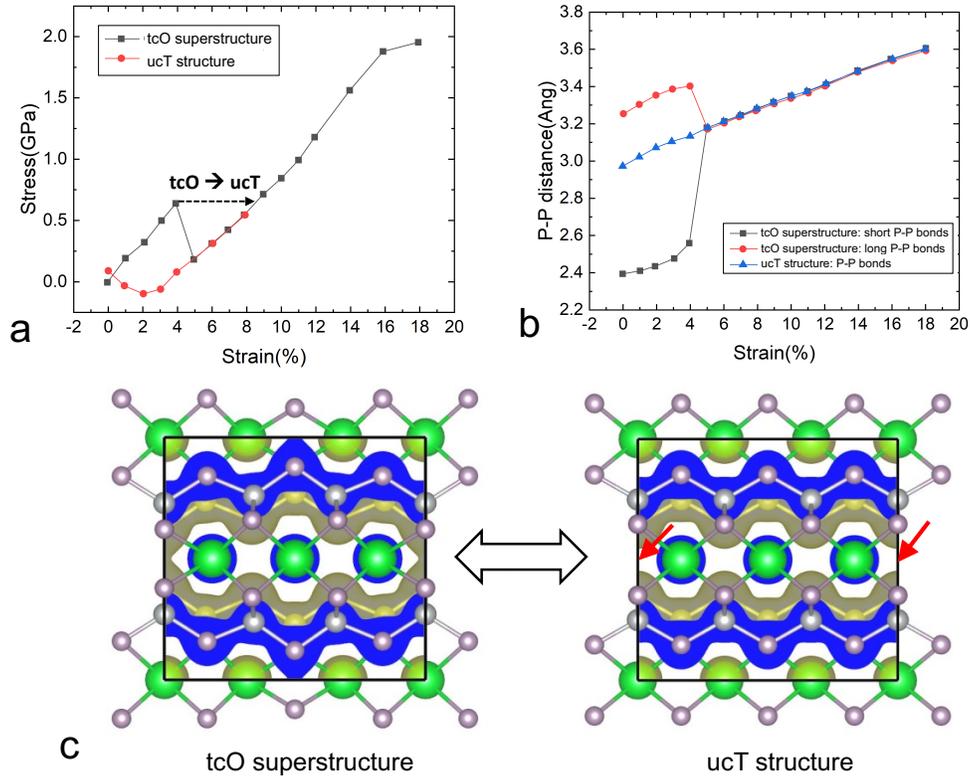

**Figure 9.** DFT calculation data of the tension test. (a) Stress vs. Strain curve. A broken line arrow indicates the transition under load control. (b) P-P distance vs. Strain curve, (c) Electron density before and after the structural transition of the tcO superstructure. Red arrows indicate the location where formation or breaking P-P bonds occurs.

During unloading, a fraction (~85%) of the ucT structure will collapse to the tcO superstructure, and then the two different structures will coexist. Multiple tensile loading-unloading cyclic tests confirmed that the cyclic stress-strain curves look nearly identical (Fig. 2b), implying that the volume fraction of the recovered tcO superstructure for each cycle is always similar. Thus, as for the case of the compression tests, the volume fraction of each phase at zero stress state seems to be determined by thermodynamics, and the co-existence of two structures appears to be energetically more stable than the presence of a single structure. The most likely explanation for the phase separation in $SrNi_2P_2$ is the effects of configurational entropy. Having two different structures increases the configurational entropy, which could exceed the enthalpy difference between the two structures, leading to a lower Gibbs free energy for the two-phase mixture. Under the application of a mechanical force, however, this equilibrium will be shifted. For instance, $SrNi_2P_2$ will tend to have more bonded P atoms (the tcO superstructure) under compressive forces because making P-P bonds could relax the strain energy under compression. Conversely, $SrNi_2P_2$ will tend to have more unbonded P atoms (the ucT structure) under tensile forces.

The combined stress-strain curve from the compression and tension tests (Fig. 10a) shows that there are a total 16 stages of recoverable deformation. This unique superelastic process is possible due to the presence of two different structures and their unique lattice collapse and expansion mechanism. Table 1 summarizes the deformation mode of the structural transition in each stage. Note that most $ThCr_2Si_2$-type intermetallic compounds have either the ucT structure or the cT structure. They could exhibit either compressive superelasticity by forming Si-Si type bonds in the ucT structure or tensile superelasticity by breaking Si-Si type bonds in the cT structure. However, it is impossible to observe both compressive and tensile superelasticity from a single-phase $ThCr_2Si_2$-type intermetallic compound, and the existence of compressive and tensile superelasticity in $SrNi_2P_2$ is an unusual case, allowing for a wide range of superelastic strain (20%) when compression and tension are considered together (Fig. 10b).

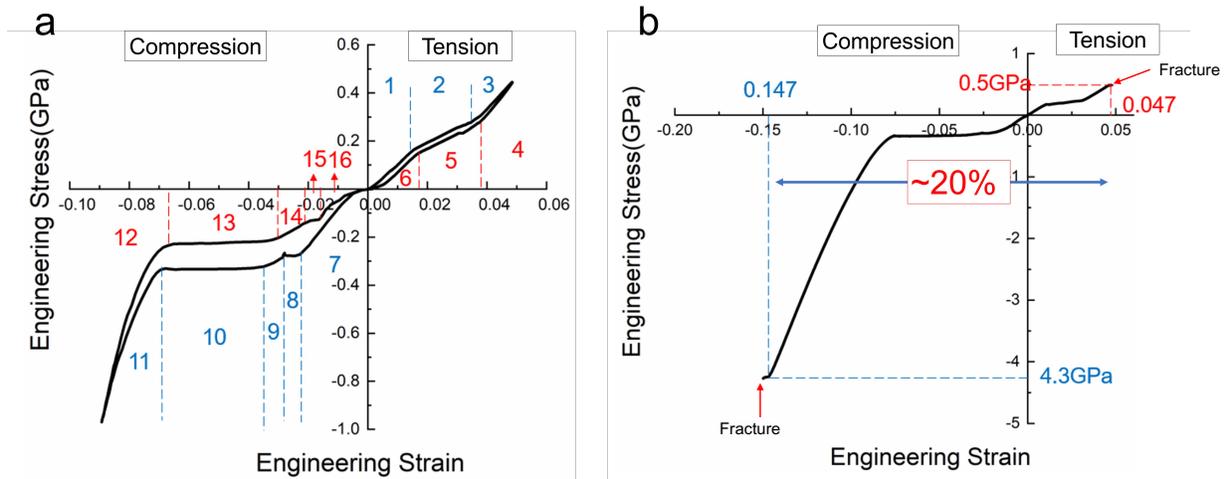

**Figure 10.** Combined stress-strain curves of compression and tension tests. (a) Cyclic test (b) Loading until facture. Note that to combine compression and tension stress-strain curves in a single plot, compressive stresses are presented with negative values, and tensile stresses are presented with positive values by convention.

To observe the structural transition process directly, in-situ TEM or in-situ XRD measurements could be useful to obtain direct experimental evidence of the structural transition processes [33]. However, all the experimental data (XRD, TEM, mechanical data and DFT calculations) strongly suggest that our analysis provides a highly reasonable explanation for the structural transition process under compression and tension in terms of making and breaking P-P bonds, which is a well-known structural transition process in $ThCr_2Si_2$-type intermetallic compounds. Based on the geometry of the atomic configuration of $SrNi_2P_2$, the structural transition should occur along the direction of P-P bond formation under compression (the ucT structure with no P-P bonds → the tcO superstructure with P-P bonds between one third of P atoms → the cT structure with P-P bonds between all P atoms) and should occur along the direction of P-P bond breakage under tension (the tcO superstructure → the ucT structure).

| # | Process | | Microstructure |
|---|---|---|---|
| 1 | Tension loading | Elastic tension | tcO (85%) + ucT (15%) |
| 2 | | Phase transition | tcO (85%) + ucT (15%) → ucT (100%) |
| 3 | | Elastic tension | ucT (100%) |
| 4 | Tension unloading | Elastic recovery | ucT (100%) |
| 5 | | Phase transition | ucT (100%) → tcO (85%) + ucT (15%) |
| 6 | | Elastic recovery | tcO (85%) + ucT (15%) |
| 7 | Compression loading | Elastic compression | tcO (85%) + ucT (15%) |
| 8 | | Phase transition | tcO (85%) + ucT (15%) → tcO (100%) |
| 9 | | Elastic compression | tcO (100%) |
| 10 | | Phase transition | tcO (100%) → cT (100%) |
| 11 | | Elastic compression | cT (100%) |
| 12 | Compression unloading | Elastic recovery | cT (100%) |
| 13 | | Phase transition | cT (100%) → tcO (100%) |
| 14 | | Elastic recovery | tcO (100%) |
| 15 | | Phase transition | tcO (100%) → tcO (85%) + ucT (15%) |
| 16 | | Elastic recovery | tcO (85%) + ucT (15%) |

**Table 1.** Deformation mode and microstructure in each stage during compression and tension.

### 4.3 Temperature effect on the structural transition

Based on Hoffman and Zhang, the distance between Si-type atoms in $ThCr_2Si_2$-type intermetallic compounds is one of the most important factors that determine the bonding state of Si-type atoms (in our case, phosphorous) [34]. As the distance of Si-type atoms is reduced, it is easier to form the Si-Si bonds. In the extreme case, if Si-type atoms are too close to each other, the Si-Si bonds are formed even without any applied stress, and the collapsed state could be the energetically stable structure. Under compression, the first plateau region in the stress-strain data of $SrNi_2P_2$ corresponds to the structural transition from the ucT structure, which occupies 15% of the entire volume, to the tcO superstructure. The experimental data showed that as the temperature decreases, the critical stress of the first transition decreases. As described in the previous section, the ucT structure is under a small residual compressive stress, which reduces the P-P distance. At lower temperatures, thermal contraction could reduce the P-P distance even further. Then, a smaller applied stress would be sufficient enough to form P-P bonds and to induce the structural transition from the ucT structure to the tcO superstructure by forming P-P bonds for one third of P bonds. The same reasoning can be applied to the lower critical stress for the second transition at lower temperatures. The P-P distance between unbonded P atoms in the tcO superstructure would also be shorter at lower temperature due to the thermal contraction. Thus, at lower temperatures, a smaller stress is required to induce the structural transition from the tcO superstructure to the cT structure. A similar temperature dependence was observed in $CaFe_2As_2$, which also showed a lower stress for the structural transition at lower temperatures [35].

Under tension, the critical stress of the structural transition increases as the temperature decreases. At lower temperatures, the P-P distance of bonded P atoms is shorter, implying a tighter bonding. Thus, a higher stress is required to break P-P bonds in the tcO superstructure. Additionally, at lower temperatures, thermal vibrations (which tend to break P-P bonds) are weak, so it is necessary to apply a higher stress to break P-P bonds. Once the critical stress becomes greater than the fracture strength, a specimen could fracture before any structural transition occurs, leaving no sign of superelasticity in tension stress-strain data. We can observe this trend in the data acquired at both 200K (fracture right after the transition ends) and 100K (fracture in the middle of the transition).

Note that the temperature dependence of the critical stress for the structural transition is quite different from those for conventional SMAs. Most SMAs show superelasticity via a reversible martensitic phase transformation. This same shear transformation occurs regardless of the direction of the shear stress. Thus, the critical stress of the structural transition always decreases as the temperature decreases regardless of the direction of the shear stress. In the case of $SrNi_2P_2$, however, the structural transition under compression occurs via the formation of P-P bonds, but the structural transition under tension occurs via the breakage of P-P bonds. Thus, opposite processes control the structural transition, leading to the opposite temperature effects on the critical stress; *i.e.*, the critical stress of the structural transition decreases under compression but increases under tension.

Based on all the experimental data, a temperature-stress phase diagram of $SrNi_2P_2$ was constructed (Fig.11). The x-axis in this diagram is uniaxial stress and the y-axis is the absolute temperature. The boundaries between microstructural states were determined from our micro-compression and micro-tensile tension tests under room and cryogenic temperatures as well as from previous data of the resistance measurement. There is a shift between equivalent boundaries during loading and unloading, which represents the presence of hysteresis caused by the different stress levels required to form and break P-P bonds. The shift remains nearly the same no matter how temperature changes. At the current stage, it is difficult to describe the fundamental mechanism and it requires a future in-depth theoretical study. The extrapolated value of the critical transition stress at 0 K during loading shows a relatively good agreement with the value obtained from DFT.

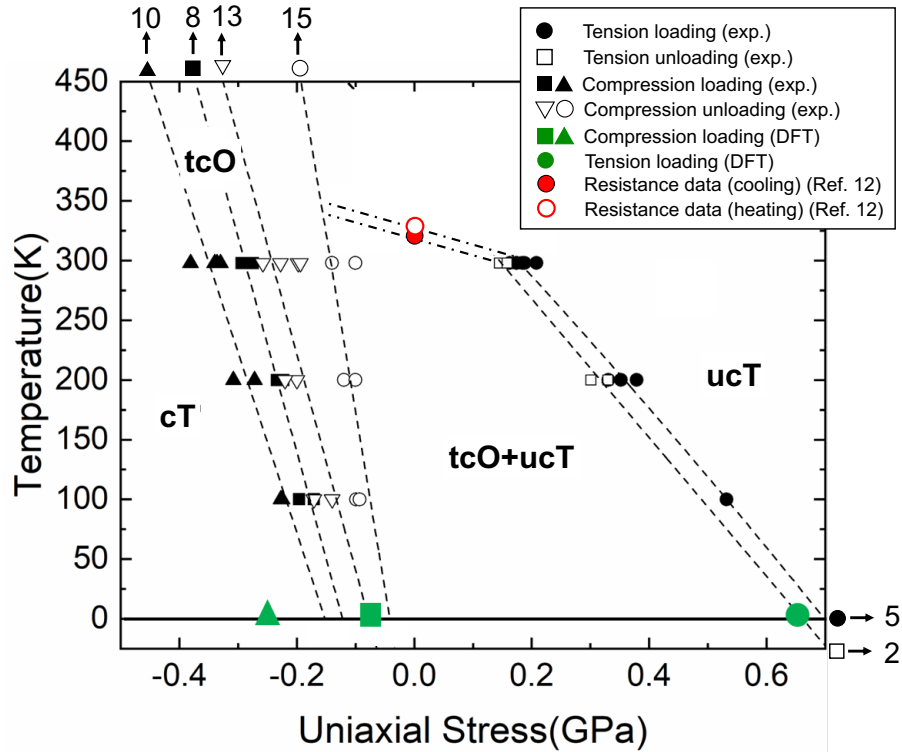

**Figure 11.** The temperature-uniaxial stress diagram of $SrNi_2P_2$. A broken line is the fitting line of each data set that corresponds to the same symbol outside of the graph. A number, which matches the symbol outside of the graph, corresponds to the process number (#) in Table 1.

### 4.4 Potential elastocaloric effect at low temperatures

Elastocaloric cooling, also called thermoelastic cooling, has been seen as the most viable alternative to conventional vapor compression cooling systems [36]. It is a cooling effect that occurs when SMAs undergo a reversible transformation under an applied stress. For conventional SMAs, elastocaloric cooling is related to the latent heat associated with the martensite phase transformation process. When the austenite phase is transformed into the martensite phase under loading, the entropy is reduced, and the latent heat is released. For the reverse phase transformation from the martensite to the austenite phases during unloading, on the other hand, the entropy is increased, and the SMAs absorb heat from their surroundings, causing a decrease in the temperature of the environment. The Clausius-Clapeyron equation [14] is the constitutive relation of elastocaloric materials with the latent heat of the phase transformation:

$$\frac{d\sigma}{dT} = -\frac{\rho \Delta S}{\varepsilon_{SE}}, \tag{1}$$

where $\sigma$ is the absolute value of transformation stress, $T$ the absolute temperature, $\rho$ is the mass density, $\Delta S$ the volumetric entropy change, and $\varepsilon_{SE}$ the phase transformation strain. Combined with the relation between the latent heat ($Q$) and the adiabatic temperature, we can evaluate the potential temperature change of the environment ($\Delta T$) due to the elastocaloric effect by

$$\Delta T = \frac{Q}{c_p} = -\frac{T}{c_p} \Delta S, \tag{2}$$

where $c_p$ is specific heat.

From the stress-strain data, the transformation strain of SrNi$_2$P$_2$ is 4% for compression (the Stage II (0.5%) and Stage IV (3.5%) in Fig. 2a) and 2% for tension (the low slope region, Stage II in Fig. 2b). The mass density of SrNi$_2$P$_2$, which contains 83.5% of the tcO superstructure and 16.5% of the ucT structure [9], is calculated as 5426.21 kg/m$^3$ (See Supplementary Materials). As discussed before, the temperature dependence of the critical stress for the structural transition is opposite between compression and tension. For compressive loading, the critical compressive stress decreases as the temperature decreases with an average slope ($d\sigma/dT$ in Eq. (1)) of 0.58 MPa/K, which corresponds to $\Delta S = -5.34$ J/kg/K (Fig.12). This trend is similar to that of conventional SMAs under either compressive or tensile loading. Under tensile loading, however, the critical tensile stress for the structural transition increases as the temperature decreases with the average slope of -1.8 MPa/K, which corresponds to $\Delta S = 6.6$ J/kg/K (Fig.12). According to Eq. (1), these results mean that $\Delta S$ of $SrNi_2P_2$ changes sign if the loading direction is changed because the sign of $d\sigma/dT$ changes. In other words, $\Delta S$ of $SrNi_2P_2$ maintains the same sign if the loading direction does not change. This result is opposite to the trend in conventional SMAs, where the sign of $d\sigma/dT$ or $\Delta S$ is switched if the sign of stress is changed even though the loading direction is unchanged.

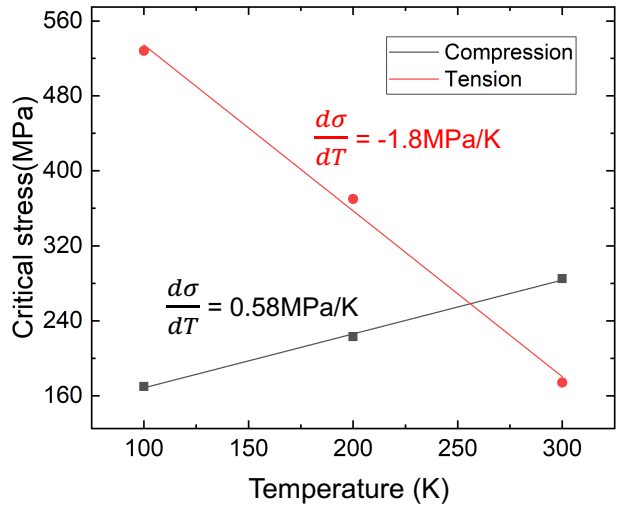

**Figure 12.** Temperature dependence of the critical stress of structural transition in $SrNi_2P_2$

According to Eq. (2), SMAs can heat up the environment (positive $\Delta T$) during both compressive and tensile loading due to the negative $\Delta S$ and cool down (negative $\Delta T$) during both compressive and tensile unloading due to the positive $\Delta S$ (Fig. 13a). However, $SrNi_2P_2$ can heat up the environment during compressive loading or tensile unloading and can cool down during compressive unloading or tensile loading (Fig. 13b). In other words, $SrNi_2P_2$ can heat up the environment continuously through deformation from the tensile elastic strain limit to the compressive elastic strain limit or cool down the environment by deformation from the compressive elastic strain limit to the tensile elastic strain limit. Thus, the same heating or cooling process occurs throughout a large deformation from one elastic strain limit to the other. However, conventional SMAs change the sign of temperature change ($\Delta T$) once the sign of stress is changed, so the elastocaloric effect is discontinuous as the sign of applied stress is changed during the cyclic deformation (Figure 13). The continuous elastocaloric effect of SrNi$_2$P$_2$ is greatly beneficial because a single directional deformation could induce heating or cooling twice, which could lead to more effective elastocaloric heating or cooling.

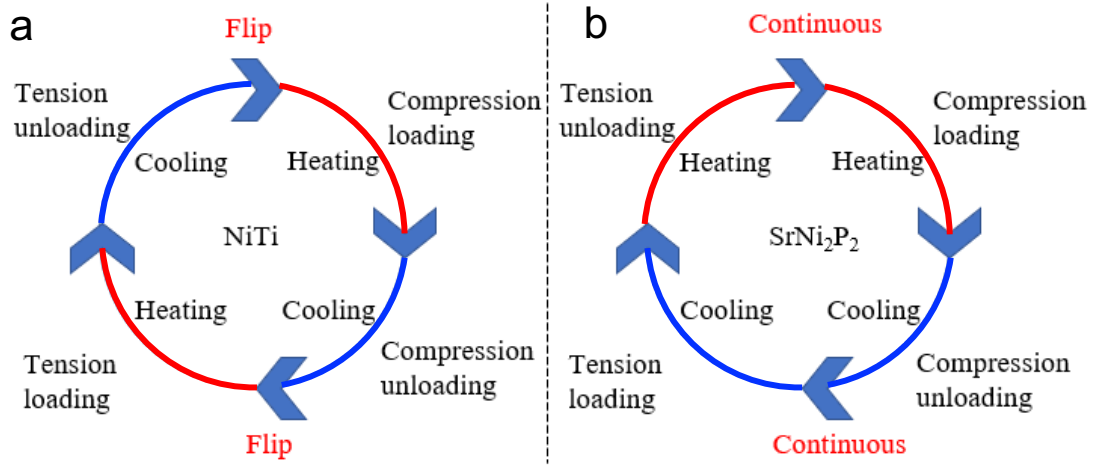

**Figure 13.** Schematic diagram of elastocaloric cooling effect during the loading and unloading process of tension and compression in (a) conventional SMAs (NiTi) (b) SrNi$_2$P$_2$

To compute the temperature change from Eq. (2), $c_p$ is required. To our knowledge, $c_p$ of SrNi$_2$P$_2$ has only been measured at low temperatures [37], but it is possible to estimate $c_p$ using the Debye model (Phonon contribution) with the Debye temperature ($\theta_D$) of SrNi$_2$P$_2$ (T$_D$=348 K [37]) and the Sommerfeld model (electron contribution) with the Sommerfeld coefficient ($\gamma$=15 mJ/mol/K [37]) from the following expression:

$$c_p \approx c_v = c_{Debye} + c_{Sommerfelt}$$

$$= 9NR\left(\frac{T}{\theta_D}\right)^3 \int_0^{\frac{\theta_D}{T}} \frac{x^4 e^x}{(e^x-1)^2} dx + \gamma T \tag{3}$$

where $N = 5$. The combination of these two models usually works very well for most materials [38].

Based on our calculation, $c_p$ is 74.32, 110, and 121.21J/mol/K at 100K, 200K, and RT, respectively. Note that at RT (300K), the $c_p$ value is already close to the Dulong-Petit limit (124.71 J/mol/K). Eq. (2) gives the temperature changes for the elastocaloric cooling (compression unloading and tension loading), which are summarized in Table 2. For a continuous deformation in one direction from compression unloading to tension loading, it is possible to obtain a temperature change of -12.86, -8.63K, and -7.89K at 100K, 200K, and RT, respectively. Although the individual loading section (either $\Delta T_{cu}$ or $\Delta T_{tl}$; $cu$ = compressive unloading, $tl$ = tension loading) does not show an outstanding elastocaloric cooling, their sum ($\Delta T_{cu} + \Delta T_{tl}$) is comparable with those of high performance SMAs. Therefore, the asymmetric superelastic process of SrNi$_2$P$_2$ eventually leads to an excellent elastocaloric effect.

Unfortunately, in this study, we were not able to measure the temperature change. This is partly because the micropillars are too small to attach temperature sensors. Moreover, the temperature of such small micropillars could be quickly equilibrated with the surroundings, which is a giant heat bath. It would be necessary to grow a single crystal large enough to overcome these two issues. Although the temperature measurement is difficult under the current situation, the cryogenic micromechanical testing allowed us to obtain $\Delta S$. Also, the combination of Debye and Sommerfeld models captures the temperature-dependent specific heat quite precisely in most cases. Therefore, we strongly believe that our estimation of the temperature change is quite accurate.

High spatial resolution spectroscopy or MEMS temperature sensors could be useful to measure the temperature change directly.

| $T$ | $c_p$ | Compressive unloading ($\Delta T_{cu}$) | Tension loading ($\Delta T_{tl}$) | $\Delta T_{cu} + \Delta T_{tl}$ |
|---|---|---|---|---|
| 100 | 74.32 | -4.31 | -7.11 | -11.42 |
| 200 | 110 | -2.90 | -4.77 | -7.67 |
| 300 | 121.21 | -2.65 | -4.36 | -7.01 |

**Table 2.** Summary of $c_p$ and the temperature change under compressive unloading and tension loading. Note that the temperature change under compressive loading and tension unloading will be similar with that under compressive unloading and tension loading because $d\sigma/dT$ of our experimental data is similar regardless of loading or unloading for both compression and tension.

## 5. Concluding Remarks

The superelastic behavior and the related structural transitions of SrNi$_2$P$_2$ micropillars were investigated for tension and compression at different low temperatures using *in-situ* micro-mechanical testing, transmission electron microscopy, and density functional theory calculations. Under compression, a double lattice collapse occurs due to an initial transition from the ucT structure to the tcO superstructure and then a second transition from the tcO superstructure to the cT structure, and these two transitions are reversed upon unloading. Notably, unlike most ThCr$_2$Si$_2$-structured intermetallic compounds, SrNi$_2$P$_2$ also exhibits superelasticity even under tension via a single lattice expansion from the tcO superstructure to the ucT structure, and this transition can be reversed upon unloading. The simultaneous presence of both tension and compression superelasticity results from the presence of the tcO superstructure under the zero-stress state. Due to the different transition mechanisms, the shape of the compressive stress-strain curve is entirely different from that of the tensile stress-strain curve. Also, the temperature dependence of the stress-strain data is opposite between compression and tension. As the temperature decreases, the critical transition stress decreases under compression, but increases under tension. DFT simulation results indicate that the tension-compression asymmetry in stress-strain response and its temperature dependence results from the opposite structural transition mechanisms, forming P-P bonds under compression and breaking P-P bonds under tension. Interestingly, this opposite structural transition process could lead to unique elastocaloric cooling effects. Even though the sign of stress is changed, cooling or heating can be done continuously if the direction of loading is unchanged, for instance, from the tensile elastic limit to the compressive elastic limit or vice versa. This continuous elastocaloric effect leads to ~10K temperature change, which is similar to the performance of SMAs. Our work unveiled the unique superelasticity mechanisms of $SrNi_2P_2$ under compression and tension and confirmed the remarkably large recoverable strain and a continuous elastocaloric cooling effect. The excellent performance of $SrNi_2P_2$ holds great promise for its potential structural and aerospace applications such as cryogenic actuators, heating and cooling elements used in cryogenic environments.

**Acknowledgements**

SX, and S-WL are supported by the Early Career Faculty Grant from NASA's Space Technology Research Grants Program (NNX16AR60G). The FIB milling and TEM studies were performed using the facilities in the UConn/Thermo Fisher Scientific Center for Advanced Microscopy and Materials Analysis (CAMMA). PCC is supported by the U.S. Department of Energy, Office of Basic Energy Science, Division of Materials Sciences and Engineering. Ames Laboratory is operated for the U.S. Department of Energy by Iowa State University under Contract No. DE-AC02-07CH11358. VB acknowledges the computational resources provided by the computer center of Goethe University Frankfurt. AV and RV acknowledges the support by the Deutsche Forschungsgemeinschaft (DFG, German Re-search Foundation) for funding through TRR 288 – 422213477 (Project A05). We also thank Guilherme Gorgen-Lesseux for initial crystal growth work.